\documentclass[vecphys]{svmult}
\usepackage{amsmath,amssymb,amscd}
\usepackage{makeidx}
\usepackage{multicol}
\usepackage[bottom]{footmisc}

\makeindex

\begin{document}

\title*{Quantum Field Theory on Curved Backgrounds}

\author{Romeo Brunetti\inst{1}\and Klaus Fredenhagen\inst{2}}
\institute{Dipartimento di Matematica, Universit\`a di Trento, 
\texttt{brunetti@science.unitn.it}
\and
II Inst. f. Theor. Physik, Universit\"at Hamburg,
\texttt{klaus.fredenhagen@desy.de}}

\maketitle

\section{Introduction}\label{K_intro}
Quantum field theory is an extremely successful piece of theoretical physics. Based on few general principles, it describes with an incredibly good precision large parts of particle physics. But also in other fields, in particular in solid state physics, it yields important applications. At present, the only problem which seems to go beyond the general framework of quantum field theory is the incorporation of gravity. Quantum field theory on curved backgrounds aims at a step towards solving this problem by neglecting the back reaction of the quantum fields on the spacetime metric.

Quantum field theory has a rich and rather complex structure. It appears in different versions which are known to be essentially equivalent. Unfortunately, large parts of the theory are available only at the level of formal perturbation theory, and a comparison of the  theory with experiments requires a truncation of the series which is done with a certain arbitrariness.

Due to its rich structure, quantum field theory is intimately related to various fields of mathematics and has often challenged the developments of new mathematical concepts.

In these lectures, we will give an introduction to quantum field theory in a formulation which admits a construction on generic spacetimes. Such a construction is possible in the so-called algebraic approach to quantum field theory \cite{K-HK,K-Haag}. The more standard formulation as one may find it in typical text books (see, e.g.,  \cite{K-Peskin}) relies heavily on concepts like vacuum, particles , energy and makes strong use of the connection to statistical mechanics via the so-called Wick rotation. But these concepts loose their meaning on generic Lorentzian spacetimes and are therefore restricted to a few examples with high symmetry. It was a major progress of recent years that local versions of most of these concepts have been found. Their formulation requires the algebraic framework of quantum physics and, on the more technical side, the replacement of momentum space techniques by techniques from microlocal analysis.

The plan of the lectures is as follows. After a general discussion of fundamental physical concepts like states, observables and subsystems we will describe a general framework that can be used to
define both classical and quantum field theories. It is based on the locally covariant approach to quantum field theory \cite{K-BFV} which uses the language of categories to incorporate the principle of general covariance.

The first example of the general framework is 
the canonical formalism of classical field theory based on the so-called Peierls bracket by which the algebra of functionals of classical field configurations is endowed with a Poisson structure.

We then present as a simple example in quantum field theory the free scalar quantum field.

A less simple example is the algebra of Wick polynomials of the free field. Here, for the first time, techniques from microlocal analysis enter. The construction relies on a groundbreaking observation of Radzikowski \cite{K-Radman}. Radzikowski found that the so-called Hadamard condition on the 2-point correlation function is equivalent to a positivity condition on the wave front set, whose range of application was extended and named ``microlocal spectrum condition'' few years later \cite{K-BFK}.   
This insight not only, for the first time, permitted the construction of nonlinear fields on generic spacetimes, but also paved the way for a purely algebraic construction, which before was also unknown on Minkowski space. 

Based on these results, one now can construct also interacting quantum field theories in the sense of formal power series. The construction can be reduced to the definition of time ordered products of prospective Lagrangians. By the principle of causality, the time ordered products of $n$ factors are determined by products (in the sense of the algebra of Wick polynomials) of time ordered products of less than $n$ factors outside of the thin diagonal $\Delta_n\subset{\mathcal{M}}^n$ (considered as algebra valued distributions). The removal of ultraviolet divergences amounts in this framework to the extension of distributions on ${\mathcal{M}}^n\setminus \Delta_n$ to ${\mathcal{M}}^n$. The possible extensions can be discussed in terms of the so-called microlocal scaling degree which measures the singularity of the distribution transversal to the submanifold $\Delta_n$.
   
\section{Systems and subsystems}\label{K_ssub}
\subsection{Observables and states}\label{K_obst}
Experiments on a physical system may be schematically described as maps
\begin{equation}
\mathrm{experiment}:(\mathrm{state},\mathrm{observable})\mapsto \mathrm{result} \ .
\end{equation}
Here a state is understood as a prescription for the preparation of the system, and the observable is an operation on the prepared system which yields a definite result. In classical physics, one assumes that an optimally prepared system (``pure state") yields for a given (ideal) observable always the same result (which may be recorded as a real number). Thus observables can be identified with real valued functions on the set of pure states. The set of observables so gets the structure of an associative, commutative algebra over ${\mathbb{R}}$, and the pure states are reobtained as characters of the algebra, i.e. homomorphisms into ${\mathbb{R}}$.

In classical statistical mechanics one considers also incomplete preparation prescriptions, e.g. one puts a number of particles into a box with a definite total energy, but without fixing positions and momenta of the individual particles. Such a state corresponds to a probability measure $\mu$ on the set of pure states, or, equivalently, to a linear functional on the algebra of observables which is positive on positive functions and assumes the value 1 on the unit observable. For the observable $f$ the state yields the probability distribution  
\begin{equation}
(\mu,f)\mapsto f_{\star}\mu \ ,\ f_{\star}\mu(I)=\mu(f^{-1}(I)) 
\end{equation}
on ${\mathbb{R}}$. Pure states are the Dirac measures. 

In quantum mechanics, the measurement results fluctuate even in optimally prepared states. Pure states are represented by 1-dimensional subspaces $\mathfrak{L}$ of some complex Hilbert space, and observables are identified with selfadjoint operators $A$. The probability distribution of measured values is given by
\begin{equation}
\mu_{A,\mathfrak{L}}(I)=(\Psi,E_A(I)\Psi)
\end{equation}
where $\Psi$ is any unit vector in $\mathfrak{L}$ and $E_A(I)$ is the spectral projection of $A$ corresponding to the interval $I$.

In quantum statistics, one admits a larger class of states, corresponding to incomplete preparation, which can be described by a density matrix, i.e. a positive trace class operator $\rho$ with trace 1; the probability distribution is given by
\begin{equation}
\mu_{A,\rho}(I)=\mathrm{Tr}\rho E_A(I) \ 
\end{equation}
where the pure states correspond to the rank one density matrices.

In spite of the apparently rather different structures one can arrive at a unified description. The set of observables is a real vector space with two products:\begin{enumerate}
\item a commutative, but in general nonassociative product (the Jordan product),
\begin{equation}
A\circ B=\frac14 \left((A+B)^2-(A-B)^2\right) \ ,
\end{equation}
arising from the freedom of relabeling measurement results;

\item an antisymmetric product
\begin{equation}
\{A,B\}
\end{equation}
which is known as the Poisson bracket in classical mechanics and is given by $\frac{i}{\hbar}$ times the commutator $[\cdot,\cdot] $ in quantum mechanics. This product originates from the fact that every observable $H$ can induce a transformation of the system by Hamilton's    (or Heisenberg's) equation
\begin{equation}
\frac{d}{dt}A(t)=\{H,A(t)\} \ .
\end{equation}

\end{enumerate}
The two products satisfy the following conditions:
\begin{enumerate}
\item $A\mapsto\{B,A\}$ is a derivation with respect to both products.
\item The associators of both products are related by
\begin{equation}
(A\circ B)\circ C-A\circ(B\circ C)=\frac{\hbar^2}{4}\left(\left\{\{A,B\},C\right\}-\left\{A,\{B,C\}\right\}\right) \ .
\end{equation}

\end{enumerate}
While the first condition is motivated by the interpretation of Hamilton's equation as an infinitesimal symmetry, there seems to be no physical motivation for the second condition. But mathematically, it has a strong impact: in classical physics $\hbar=0$, hence the Jordan product is associative; in quantum physics, the condition implies that
\begin{equation}
AB:=A\circ B+\frac{\hbar}{2i}\{A,B\}
\end{equation}
is an associative product on the complexification ${\mathfrak{A}}={\mathfrak{A}}_{{\mathbb{R}}}\otimes {\mathbb{C}}$, where the information on the real subspace is encoded in the $\star$-operation
\begin{equation}
(A\otimes z)^*=A\otimes \overline{z} \ .
\end{equation}
States are defined as linear functionals on the algebra which assume positive values on positive observables and are 1 on the unit observable. A priori, in the case $\hbar\not=0$ the positivity condition on the subspace  ${\mathfrak{A}}_{{\mathbb{R}}}$ of selfadjoint elements could be weaker than the positivity requirement on the complexification ${\mathfrak{A}}$.
Namely, on the real subspace we call positive every square of a self adjoint element, whereas on the full algebra positive elements are absolute squares of the form
\begin{equation}
(A-iB)(A+iB)=A^2+B^2+i\hbar\{A,B\} \ , \ A,B \text{ selfadjoint }
\end{equation}
But under suitable completeness assumptions, in particular when ${\mathfrak{A}}$ is a C*-algebra, operators as above admit a selfadjoint square root, thus the positivity conditions
coincide in these cases. If one is in a more general situation, one has to require that states satisfy the stronger positivity condition, in order to insure the existence of the GNS representation.  
\subsection{Subsystems}\label{K_subsystems}
A system may be identified with a unital C*-algebra ${\mathfrak{A}}$. 
Subsystems then correspond to sub-C*-algebras $\mathfrak{B}$ with the same unit. 
A state of a system then induces a state on the subsystem by restricting the linear functional $\omega$ on ${\mathfrak{A}}$ to the subalgebra $\mathfrak{B}$. The induced state may be mixed even if the original state was pure.

One may also ask whether every state on the subalgebra $\mathfrak{B}$ arises as a restriction of a state on ${\mathfrak{A}}$. This is actually true. Namely, let $\omega$ be a state on $\mathfrak{B}$. According to the Hahn-Banach theorem, $\omega$ has an extension to a linear functional $\tilde{\omega}$ on ${\mathfrak{A}}$ with $||\tilde{\omega}||=||\omega||$. But $\tilde{\omega}(1)=\omega(1)=||\omega||=1$, hence $\tilde{\omega}$ is a state.

Two subsystems $\mathfrak{B}_1$ and $\mathfrak{B}_2$ may be called independent whenever the algebras $\mathfrak{B}_1$ and $\mathfrak{B}_2$ commute and 
\begin{equation}
B_1\otimes B_2\mapsto B_1B_2
\end{equation}
defines an isomorphism from the tensor product $\mathfrak{B}_1\otimes\mathfrak{B}_2$ to the algebra generated by $\mathfrak{B}_1$ and $\mathfrak{B}_2$.

Given states $\omega_i$ on $\mathfrak{B}_i$, $i=1,2$, one may define a product state on $\mathfrak{B}_1\otimes\mathfrak{B}_2$ by
\begin{equation}
(\omega_1\otimes\omega_2)(B_1\otimes B_2)=\omega_1(B_1)\omega(B_2) \ .
\end{equation}
Convex combinations of product states are called separable. As was first observed by Bell, there exist nonseparable states if both algebras contain subalgebras isomorphic to $\mathrm{M}_2({\mathbb{C}})$. This is the famous phenomenon of entanglement which shows that states in quantum physics may exhibit correlations between independent systems which cannot be described in terms of states of the individual systems. This is the reason, why the notion of locality is much more evident on the level of observables than on the level of states.
\subsection{Algebras of unbounded operators}\label{K_algebras}
In applications often the algebra of observables cannot be equipped with a norm. The CCR algebra is a prominent example. In these cases one usually still has a unital $*$-algebra, and states can be defined as positive normalized functionals. The GNS construction remains possible, but does not lead to a representation by bounded Hilbert space operators. In particular it is not guaranted that selfadjoint elements of the algebra are represented by selfadjoint Hilbert space operators. There is no general theory available which yields a satisfactory physical interpretation in this situation. One therefore should understand it as an intermediary step towards a formulation in terms of C*-algebras.   
\section{Locally Covariant Theories}\label{K_LCT}
\subsection{Axioms of locally covariant theories}\label{K_ALCT}
Before constructing examples of classical and quantum field theories we want to describe the minimal requirements that such theories should satisfy \cite{K-BFV}:
\begin{enumerate}
\item To each globally hyperbolic time oriented spacetime ${\mathcal{M}}$ we associate a unital $*$-algebra ${\mathfrak{A}}({\mathcal{M}})$.
\item Let $\chi:{\mathcal{M}}\to{\mathcal{N}}$ be an isometric causality preserving embedding. Then there is an injective homorphism
\begin{equation}
\alpha_{\chi}:{\mathfrak{A}}({\mathcal{M}})\to{\mathfrak{A}}({\mathcal{N}}) \ .
\end{equation}
\item Let $\chi:{\mathcal{M}}\to{\mathcal{N}}$ and $\chi':{\mathcal{N}}\to\mathcal{L}$ be admissible embeddings. Then
\begin{equation}
\alpha_{\chi\circ\chi'}=\alpha_\chi \alpha_{\chi'} \ .
\end{equation}
\end{enumerate}
\noindent These axioms characterize a theory as a covariant functor ${\mathfrak{A}}$ from the category $\mathfrak{Man}$ of globally hyperbolic time oriented Lorentzian manifolds with isometric causality preserving mappings as morphisms to the category of unital $*$-algebras $\mathfrak{Alg}$ with injective homomorphisms as morphisms. It is clear that by this we characterize abstractly the notion of subsystems as defined in the previous section.

In addition we require
\begin{enumerate}
\item[$4.$] Let $\chi_i:{\mathcal{M}}_i\to{\mathcal{N}}$, $i=1,2$, be morphisms with causally disjoint closed images. Then the images of ${\mathfrak{A}}({\mathcal{M}}_1)$ and  ${\mathfrak{A}}({\mathcal{M}}_2)$ commute.
(Einstein causality)
\item[$5.$] Let $\chi:{\mathcal{M}}\to{\mathcal{N}}$ be a morphism such that its image contains a Cauchy surface of ${\mathcal{N}}$. Then $\alpha_{\chi}$ is an isomorphism. (Time slice axiom) 
 \end{enumerate}
  
Axiom 4 is equivalent to a tensor structure. Namely, $\mathfrak{Man}$ is a tensor category by the disjoint union, $\mathfrak{Alg}$ has the tensor product as a tensor structure. Since admissible embeddings of a disjoint union map the components  into causally disjoint subregions, the  tensor property of the functor corresponds to Einstein causality. Hence the notion of independence for subsystems finds here the most general formulation. 

Axiom 5 relates to cobordisms of Lorentzian manifolds. Namely, we may associate to a Cauchy surface $\Sigma\subset{\mathcal{M}}$ the inverse limit of algebras ${\mathfrak{A}}({\mathcal{N}})$, $\Sigma\subset{\mathcal{N}}\subset{\mathcal{M}}$. The algebra ${\mathfrak{A}}(\Sigma)$ obtained in this way depends on the germ of $\Sigma$ in ${\mathcal{M}}$. The propagation from $\Sigma$ to another Cauchy surface $\Sigma'$ is described by the isomorphism
\begin{equation}
\alpha_{\Sigma'\Sigma}=\alpha_{{\mathcal{M}}\Sigma'}^{-1}\alpha_{{\mathcal{M}}\Sigma} \ .
\end{equation}

One may choose not to require Axiom 5, in which case the setting will be termed as off-shell. This is particularly fruitful in the case of perturbative quantum field theory.

\subsection{Fields as natural transformations}\label{K_fields}
The locally covariant framework offers the possibility for a new concept of fields. 
Namely, fields may be defined as natural transformations between a functor, say ${\mathcal{D}}$, that associates to each spacetime ${\mathcal{M}}$ a space of test functions ${\mathcal{D}}({\mathcal{M}})$, and the previous functor of a specific locally covariant theory. If we call it by $\Phi$, then it associates to any isometric embedding $\chi: M \longrightarrow N$ the following commutative diagram
$$\begin{CD}
{\mathcal{D}}({\mathcal{M}}) @>\Phi_{\mathcal{M}}>> {\mathfrak{A}}({\mathcal{M}})\\
@V\chi_\ast VV  @VV\alpha_\chi V\\
{\mathcal{D}}({\mathcal{N}}) @>\Phi_{\mathcal{N}}>>{\mathfrak{A}}({\mathcal{N}})
\end{CD}$$
where $\chi_\ast$ is the push forward on test function. The commutativity ensures that the field $\Phi\equiv(\Phi_{\mathcal{M}})_{{\mathcal{M}}\in\mathfrak{Man}}$ has a covariance property, namely, 
$$
\alpha_\chi \circ \Phi_{\mathcal{M}} = \Phi_{\mathcal{N}} \circ \chi_\ast\ .
$$

\section{Classical field theory}\label{K_cft}
\subsection{Classical observables}  \label{K_claobs}
Let ${\varphi}$ be a scalar field on a globally hyperbolic spacetime ${\mathcal{M}}$. The space of smooth field configurations is denoted by ${\mathfrak{C}}({\mathcal{M}}):=\mathcal{C}^{\infty}({\mathcal{M}})$. The basic observables are the evaluation functionals
\begin{equation}
{\varphi}(x)(h)=h(x), \ h\in{\mathfrak{C}}({\mathcal{M}}) \ .
\end{equation}
But pointlike fields tend to be singular objects even in classical field theory, therefore we consider \cite{K-BDF1} as our observables functionals $F:{\mathfrak{C}}({\mathcal{M}})\to{\mathbb{C}}$ which are differentiable in the sense that for every ${\varphi},h\in{\mathfrak{C}}({\mathcal{M}})$ the function $\lambda\mapsto F({\varphi}+\lambda h)$ is infinitely often differentiable and the $n$th derivative at $\lambda=0$ is for every ${\varphi}$ a symmetric distribution $F^{(n)}({\varphi})$ on ${\mathcal{M}}^n$ with compact support, such that      
\begin{equation}
\frac{d^n}{d\lambda^n}F({\varphi}+\lambda h)|_{\lambda=0}=\langle F^{(n)}({\varphi}),h^{\otimes n} \rangle\ .
\end{equation}
Moreover, $F^{(n)}$, as a map on ${\mathfrak{C}}({\mathcal{M}})\times \mathcal{C}^{\infty}({\mathcal{M}}^n)$ is continuous (see \cite{K-Hamilton} for an introduction to this mathematical notions). 

We associate to each differentiable functional $F$ the set ${\mathrm{supp}}(F)$ defined as the  closure of the union of supports of $F^{(1)}({\varphi})$ for all ${\varphi}$ and require that also this set is compact. In addition we have to impose conditions on the wave front sets of the functional derivatives (see the contribution of A. Strohmaier \cite{K-St} for the definition of wave front sets). Here we use different options:

\begin{equation}
{\mathcal{F}}_0({\mathcal{M}})=\{F \text{ differentiable }, {\mathrm{WF}}(F^{(n)}({\varphi}))=\emptyset\} \ .
\end{equation}
An example for such an observable is 
\begin{equation*}F({\varphi})=\frac{1}{n!}\int {d\mathrm{vol}_n\,} f(x_1,\ldots,x_n){\varphi}(x_1)\cdots{\varphi}(x_n)\ ,
\end{equation*}
 with a symmetric test function $f\in\mathcal{D}({\mathcal{M}}^n)$, with  the functional derivatives
\begin{equation}
\langle F^{(k)}({\varphi}),h^{\otimes k}\rangle =\frac{1}{k!}\int {d\mathrm{vol}_n\,} f(x_1,\ldots,x_n)h(x_1)\cdots h(x_k){\varphi}(x_{k+1})\cdots{\varphi}(x_n) \ .
\end{equation}

This class unfortunately does not contain the most interesting observables, namely the local ones. We call a functional $F$ local, if all functional derivatives $F^{(n)}({\varphi})$ have support on the thin diagonal $\Delta_n:=\{(x_1,\ldots,x_n)\in{\mathcal{M}}^n, x_1=\cdots=x_n\}$. Moreover, we require that their wave front sets are transversal to the tangent bundle of the thin diagonal, considered as a subset of the tangent bundle of ${\mathcal{M}}^n$. A simple example is $F=\frac12\int {d\mathrm{vol}\,} f(x){\varphi}(x)^2$ with  a test function $f\in\mathcal{D}({\mathcal{M}})$ where the second functional derivative at the origin is
\begin{equation}
\langle F^{(2)}(0),h\rangle =\int {d\mathrm{vol}\,} f(x)h(x,x) \ .
\end{equation}


The set of local functionals is denoted by ${\mathcal{F}}_{\text{loc}}({\mathcal{M}})$. The set of local functionals contains in particular the possible interactions. It is, however, not closed under products. We therefore have to introduce a further set 
\begin{equation}
{\mathcal{F}}({\mathcal{M}})=\{F \text{ differentiable }, {\mathrm{WF}}(F^{(n)}({\varphi}))\cap (\mathcal{M}^n\times (\overline{V}_+^n\cup\overline{V}_-^n))=\emptyset \}\ .
\end{equation}
This set contains the local functionals. The condition on the wavefront sets will turn out to be crucial in quantum field theory.

It can be proved \cite{K-BF2} that a local functional $F$ is determined by a smooth function of compact support $\mathcal{L}$ on the (infinite) jet bundle over ${\mathcal{M}}$. Let $j$ be the map
\begin{equation}
j({\varphi})(x)=(x,{\varphi}(x),\nabla {\varphi}(x),\ldots) \ .
\end{equation}
Then $F({\varphi})=\int {d\mathrm{vol}\,}\mathcal{L}(j({\varphi})(x))$. 

However, without resorting to the previous result, one may also argue as follows; the dynamics is given in terms of an action, e.g. $S_0=\int {d\mathrm{vol}\,}\mathcal{L}\circ j$, with
\begin{equation}
\mathcal{L}=\frac{1}{2}(g(d{\varphi},d{\varphi})-(m^2+\xi R){\varphi}^2)+V({\varphi}) \ .
\end{equation}
But this choice violates the condition on compact support; we therefore multiply $\mathcal{L}$ by a test function $f\in\mathcal{D}({\mathcal{M}})$ which is identically to 1 in a given relatively compact region of interest ${\mathcal{N}}$ and obtain an element of ${\mathcal{F}}_{\text{loc}}({\mathcal{M}})$. We then take the Euler-Lagrange equation for the modified action $S_0$ within the region ${\mathcal{N}}$ and obtain
\begin{equation}
0=S_0^{(1)}({\varphi})=\frac{\partial\mathcal{L}}{\partial{\varphi}}-\nabla_{\mu}\frac{\partial\mathcal{L}}{\partial \nabla_{\mu}}=-(\square+m^2+\xi R){\varphi}+V'({\varphi}) \ .
\end{equation}
Since ${\mathcal{N}}$ was arbitrary the equation holds everywhere within ${\mathcal{M}}$.
\subsection{Classical M\o{}ller operators} \label{K_moller}
We now want to interpolate between different actions $S$ which differ by an element in ${\mathcal{F}}({\mathcal{M}})$, in analogy to quantum mechanical scattering theory where isometries (the famous M\o{}ller operators) are constructed which intertwine the interacting Hamiltonian, restricted to the scattering states, with the free Hamiltonian.
We interpret $S^{(1)}$ as a map from ${\mathfrak{C}}({\mathcal{M}})$ to $\mathcal{E}'({\mathcal{M}})$. We want to construct maps $r_{S_1S_2}$ (the retarded M\o{}ller operators) from ${\mathfrak{C}}({\mathcal{M}})$ to itself with the properties
\begin{eqnarray}\label{Moller}
S_1^{(1)}\circ r_{S_1S_2}&=&S_2^{(1)} \ ; \\
r_{S_1S_2}({\varphi})(x)&=&{\varphi}(x) \ , \ x\not\in J_+({\mathrm{supp}}(S_1-S_2)) \ .\label{retarded}
\end{eqnarray}
It would be interesting to know whether unique solutions exist by the Nash-Moser theorem \cite{K-BFR}. We will convince ourselves that unique solutions exist in the sense of formal power series. We set $S_1=S+\lambda F$, $S_2=S$ and differentiate (\ref{Moller}) with respect to $\lambda$. Let ${\varphi}_{\lambda}=r_{S+\lambda F,S}({\varphi})$. We obtain
\begin{equation}
\langle (S+\lambda F)^{(2)}({\varphi}_{\lambda}),\frac{d}{d\lambda}{\varphi}_{\lambda}\otimes h\rangle +\langle F^{(1)}({\varphi}_{\lambda}),h\rangle =0\ .
\end{equation}

Now we assume that the second derivatives of our actions are integral kernels of hyperbolic differential operators which possess unique retarded Green's functions $\Delta^R$ . Together with condition (\ref{retarded})
this implies that the M\o{}ller operators satisfy the differential equation
\begin{equation}
\frac{d}{d\lambda}{\varphi}_{\lambda}=-\Delta^R_{S+\lambda F}({\varphi}_{\lambda})F^{(1)}({\varphi}_{\lambda})
\end{equation}
which has a unique solution in terms of a formal power series in $\lambda$.

\subsection{Peierls bracket}\label{K_peierls}
The M\o{}ller operators can be used to endow the algebra of functionals with a Poisson bracket. This was first proposed by Peierls \cite{K-Peierls}, a complete proof was given much later by Marolf \cite{K-Marolf} (see also \cite{K-de Witt}). 

One first defines the retarded product of two functionals $F$ and $G$ by
\begin{equation}
R_S(F,G)=\frac{d}{d\lambda}G\circ r_{S+\lambda F,G}|_{\lambda =0} \ .
\end{equation}
The Peierls bracket is then a measure for the mutual influence of two possible interactions,
\begin{equation}
\{F,G\}_S=R_S(F,G)-R_S(G,F) \ .
\end{equation}
In Peierls original formulation the functionals were restricted to solutions of the Euler-Lagrange equations for $S$. It is then difficult to prove the Jacobi identity. Peierls does not give a general proof and shows instead that his bracket coincides in typical cases with the Poisson bracket in a Hamiltonian formulation.

In our off shell formalism the Peierls bracket has the form
\begin{equation}\label{Peierls}
\{F,G\}_S=\langle F^{(1)},\Delta_S G^{(1)}\rangle 
\end{equation}
with the commutator function $\Delta_S=\Delta_S^R-\Delta_S^A$.

The triple $({\mathcal{F}}({\mathcal{M}}), S, \{\cdot,\cdot\}_S)$ is termed Poisson algebra over $S$.
\\[2ex]
\subsection{Local covariance for classical field theory}\label{K_claLCT}
We want to show that classical field theory is locally covariant provided the action $S$ is a locally covariant field.

Let $\mathcal{F}$ denote the functor which associates to every $\mathcal{M}\in\mathfrak{Man}$ the commutative algebra of functionals $\mathcal{F}(\mathcal{M})$ defined before, and to every morphism $\chi:\mathcal{M}\to\mathcal{N}$ the transformation
\begin{equation}
\mathcal{F}\chi(F)(\varphi)=F(\varphi\circ\chi) \ .
\end{equation}
Since $\chi$ preserves the metric and the time orientation, forward  and backward lightcones in the cotangent bundles transform properly. Together with the covariance of the wave front sets this implies that $\mathcal{F}\chi$ maps $\mathcal{F}(\mathcal{M})$ into $\mathcal{F}(\mathcal{N})$. 

Let now $S$ be a natural transformation from $\mathcal{D}$ to $\mathcal{F}$, i.e. for every $\mathcal{M}\in\mathfrak{Man}$ we have a linear map $S_{\mathcal{M}}:\mathcal{D}(\mathcal{M})\to \mathcal{F}(\mathcal{M})$ which satisfies 
\begin{equation}
S_{\mathcal{M}}(f)(\varphi\circ\chi)=S_{\mathcal{N}}(\chi_{*}f)(\varphi) \ .
\end{equation}
Typical examples are given in terms of smooth functions $L$ of two real variables by
$S_{\mathcal{M}}(\varphi)=\int d\mathrm{vol}_{\mathcal{M}} f(x)L(\varphi(x),g_{\mathcal{M}}(d\varphi(x),d\varphi(x)))$.

We now require in addition that the second functional derivative of $S$ w.r.t. $\varphi$ is the integral kernel of a normal hyperbolic differential operator $S^{(2)}$,
i.e., for $f,h\in\mathcal{D}(\mathcal{M})$ with $f\equiv 1$ on $\mathrm{supp}(h)$ we have
\begin{equation}
\frac{d^2}{d\lambda^2}|_{\lambda=0}S_{\mathcal{M}}(f)(\varphi+\lambda h)= \int d\mathrm{vol}_{\mathcal{M}}h(x)(S_{\mathcal{M}}^{(2)}(\varphi)h(x)) \ .
\end{equation}
We then can equip $\mathcal{F}(\mathcal{M})$ with the Peierls bracket (\ref{Peierls}) and obtain a functor $\mathcal{F}_S$ from $\mathfrak{Man}$ to the category $\mathfrak{Poi}$ of Poisson algebras which satisfies the axioms 1 to 4 of locally covariant quantum field theory, where commutativity is understood as the vanishing of Poisson brackets.

\section{Quantum field theory}\label{K_QFT}
\subsection{Interpretation of locally covariant QFT}\label{K_axLCQT}
One of the main concern for the interpretation of the theory, analogous to the interpretation of quantum field theory on 
Minkowski space, is the absence of natural states. A natural state may be defined as a family of states $\omega_{{\mathcal{M}}}$ on ${\mathfrak{A}}({\mathcal{M}})$, ${\mathcal{M}}\in\mathfrak{Man}$ such that
\begin{equation}
\omega_{{\mathcal{N}}}\circ\alpha_{\chi}=\omega_{{\mathcal{M}}} \ , \ \chi:{\mathcal{M}}\to{\mathcal{N}} \ . 
\end{equation}
The ``vacuum state" used in the conventional formalism of QFT (which is also implicit in the path integral formulation) thought of as a generally covariant object may be understood as a natural state in this sense. But one can show that such a state does not exist in typical cases. This marks the most dramatic point of departure from the traditional framework of quantum field theory. The best one can do is to associate to each spacetime ${\mathcal{M}}$ a folium of states
$S({\mathcal{M}})\subset S({\mathfrak{A}}({\mathcal{M}}))$. $S$ is a contravariant functor such that
\begin{equation}
S_{\chi}\omega=\omega\circ\alpha_{\chi} \ ,\ \chi:{\mathcal{M}}\to{\mathcal{N}} \ , \ \omega\in S({\mathcal{N}}) \ .
\end{equation}

This structure allows to endow our algebras with a suitable topology, but it does not suffice for an interpretation, since it does not allow to select single states within one folium. But there is another structure which makes possible an interpretation of the theory. These are the locally covariant fields, introduced before as natural tranformations. By definition they are defined on all spacetimes simultaneously, in a coherent way. Hence states on different spacetimes can be compared in terms of their values on locally covariant fields. This can be used for instance for a thermal interpretation of states on spacetimes without a timelike Killing vector \cite{K-BOR}.  

\subsection{Free scalar field}\label{K_scalar}
The free scalar field satisfies the Klein-Gordon equation
\begin{equation}
(\square +m^2+\xi R){\varphi}=0 
\end{equation}
which is the Euler-Lagrange equation for the Lagrangian
\begin{equation}
\mathcal{L}=\frac12(g(d{\varphi},d{\varphi})-(m^2+\xi R){\varphi}^2) \ .
\end{equation}
The Klein Gordon operator $K=\square +m^2+\xi R$ possesses unique retarded and advanced propagators $\Delta^{R,A}$, since we are on globally hyperbolic spacetimes (see e.g. \cite{K-Gin}).

The corresponding functor defining the quantum theory is constructed in the following way. For each ${\mathcal{M}}$ we consider the $*$-algebra generated by a family of elements $W_{\mathcal{M}}(f)$, $f\in\mathcal{D}_{\mathbb{R}}({\mathcal{M}})$ with the relations
\begin{align}
\label{Weyl}
 W_{\mathcal{M}}(f)^{*}   &= W_{\mathcal{M}}(-f)   \\
 W_{\mathcal{M}}(f)W_{\mathcal{M}}(g) &=e^{-\frac{i}{2}\langle f,\Delta g\rangle} W_{\mathcal{M}}(f+g)\\
 W_{\mathcal{M}}(Kf) & = W_{\mathcal{M}}(0) (\equiv 1)    
\end{align} 
This algebra has a unique C*-norm and its completion is the Weyl algebra over the symplectic space $\mathcal{D}({\mathcal{M}})/\mathrm{im}K$ with the symplectic form $\langle f,\Delta g\rangle$. With $\alpha_{\chi}(W_{\mathcal{M}}(f))=W_{\mathcal{N}}(\chi_{*}f)$ one obtains a functor satisfying also the axioms 4 and 5. Moreover, $W=(W_{\mathcal{M}})$ is a locally covariant field.
It is, however, difficult to find other locally covariant fields for this functor.\\[2ex]
The free field itself is thought to be related to the Weyl algebra by the formula
\begin{equation}
W_{\mathcal{M}}(f)=e^{i{\varphi}_{\mathcal{M}}(f)} \ .
\end{equation}
This relation can be established in the so-called regular representations of the Weyl algebra, in which the one parameter groups $W_{\mathcal{M}}(\lambda f)$ are strongly continuous.
But one can also directly construct an algebra generated by the field itself. It is the unital $*$-algebra generated by the elements ${\varphi}_M(f)$, $f\in\mathcal{D}({\mathcal{M}})$ by the relations
\begin{align}
\label{CCR}
   f \mapsto                   {\varphi}_M(f) & \text{ is linear}   \\
   {\varphi}_M(f)^*=              & {\varphi}_M(\overline{f})    \\
  [{\varphi}_M(f),{\varphi}_M(g)]=& i\langle f,\Delta g\rangle \\
  {\varphi}_{\mathcal{M}}(Kf)=               & 0  \label{eom}
\end{align}
Again one obtains a functor which satisfies axioms 1-5.
If we omit the condition (\ref{eom}) (then the time slice axiom is no longer valid and one is on the off-shell formalism), the algebra may be identified with the space of functionals on the space of field configurations ${\mathfrak{C}}({\mathcal{M}})$,
\begin{equation}\label{expansion}
F({\varphi})=\sum_{\text{finite}}\int {d\mathrm{vol}_n\,} f_n(x_1,\ldots,x_n){\varphi}(x_1)\cdots {\varphi}(x_n)
\end{equation}
where $f_n$ is a finite sum of products of test functions in one variable, and where the product is given by
\begin{equation}\label{product}
(F\ast G)({\varphi})=\sum_n \frac{i^n\hbar^n}{2^n n!}\langle F^{(n)},\Delta^{\otimes n}G^{(n)}\rangle
\end{equation}
Hence, as a vector space, it may be considered as a subspace of the space ${\mathcal{F}}_0({\mathcal{M}})$ known from classical field theory. As a formal power series in $\hbar$, the product remains well defined on this larger space. 
\subsection{The algebra of Wick polynomials}\label{K_wick}
In order to include pointwise products into the formalism we have to admit more singular coefficients in the expansion (\ref{expansion}). But then the product may become ill defined. As an example consider the functionals 
\begin{align}
\label{}
   F({\varphi}) = & \int {d\mathrm{vol}\,} f(x){\varphi}(x)^2  \\
   G({\varphi}) = & \int {d\mathrm{vol}\,} g(x){\varphi}(x)^2
\end{align}
with test functions $f$ and $g$. Insertion into the formula for the product yields
\begin{equation}  \label{square}
(F\ast G)({\varphi})= \int d\mathrm{vol}_2\, f(x)g(y)\left({\varphi}^2(x){\varphi}^2(y)+4i\hbar\Delta(x,y){\varphi}(x){\varphi}(y)-2\hbar^2 \Delta(x,y)^2\right)
\end{equation}
The problematic term is the square of the distribution $\Delta$. Here the methods of microlocal analysis enter. Namely the wave front set of $\Delta$ is
\begin{align}
{\mathrm{WF}}(\Delta)=\{&(x,y;k,k'),x\text{ and }y \text{ are connected by a null geodesic }\gamma,\\
                                   & k\|g(\dot{\gamma},\cdot),U_\gamma k+k'=0, U_\gamma \text{ parallel transport along }\gamma \}
\end{align}
The product of $\Delta$ cannot be defined in terms of H\"ormander's criterion for the multiplication of distribution, since the sum of 2 vectors in the wave front set can yield zero. The crucial fact is now that $\Delta$ can be split in the form  
 \begin{equation}\label{split}
\Delta=\frac12 \Delta+iH +\frac12\Delta -iH
\end{equation}
where the "Hadamard function" $H$ is symmetric and the wave front set of $\frac12\Delta+iH$ contains only the positive frequency part
\begin{equation}
{\mathrm{WF}}(\frac12\Delta +iH)=\{(x,y;k,k')\in{\mathrm{WF}}(\Delta),k\in\overline{V_+}\} \ .
\end{equation}
On Minkowski space, $\Delta$ depends only on the difference $x-y$, and one may find $H$ in terms of the Fourier transform of $\Delta$
\begin{equation}
\frac12\Delta+iH=\Delta_+ \ , \tilde{\Delta}_+(k)=\left\{
\begin{array}{ccc}
 \tilde{\Delta}(k) & ,  & k\in\overline{V_+}  \\
             0           & ,  & \text{else}  
\end{array}
\right.
\end{equation}
On a generic spacetime, the split (\ref{split}) represents a microlocal version of the decomposition into positive and negative energies which is fundamental for quantum field theory on Minkowski space.

If we replace in the definition of the product (\ref{product}) $\Delta$ by $\Delta+2iH$, we obtain a new product $\ast_H$. On ${\mathcal{F}}_0({\mathcal{M}})$ this product is equivalent to $\ast$, namely
\begin{equation}
F\ast_H G=\alpha_H(\alpha_H^{-1}(F)\ast \alpha_H^{-1}(G))
\end{equation}
where
\begin{equation}\label{alpha_H}
\alpha_H(F)=\sum\frac{\hbar^n}{n!}\langle H^{\otimes n},F^{(2n)}\rangle
\end{equation}
is a linear isomorphism of ${\mathcal{F}}_0({\mathcal{M}})[[\hbar]]$. 

This product now  yields well defined expressions in (\ref{square}); moreover, it is well defined on ${\mathcal{F}}({\mathcal{M}})$. Up to taking the quotient by the ideal of the field equation we obtain, on Minkowski space, the algebra of Wick polynomials. We thus succeeded to define on generic spacetimes an algebra containing all local field polynomials.

The annoying feature however is that this algebra depends on the choice of $H$. Fortunately, the difference $w$ between two Hadamard functions $H$ and $H'$ is smooth.
Thus the products $\ast_H$ and $\ast_{H'}$ are equivalent,
\begin{equation}
F\ast_{H'}G=\alpha_w(\alpha_w^{-1}(F)\ast_H \alpha_w^{-1}(G))
\end{equation}
where $\alpha_w$ is defined in analogy to (\ref{alpha_H}), but is now, due to the smoothness of $w$, a well defined linear isomorphism of ${\mathcal{F}}({\mathcal{M}})[[\hbar]]$. 
    
In order to eliminate the influence of $H$ we replace our functionals by families $F=(F_H)$, labeled by Hadamard functions $H$ and satisfying the coherence condition
$\alpha_w(F_H)=F_{H+w}$. The product of two such families is defined by
\begin{equation}
(F\ast G)_H=F_H\ast_H G_H
\end{equation}
We call this algebra the algebra of quantum observables and denote it by $\mathcal{A}({\mathcal{M}})$. ${\mathcal{F}}_0({\mathcal{M}})[[\hbar]]$ equipped with the product (\ref{product}) is embedded into $\mathcal{A}({\mathcal{M}})$ by
\begin{equation}
F\mapsto (F_H) \text{ with } F_H= \alpha_H(F) \ . 
\end{equation}
One may equip ${\mathcal{F}}({\mathcal{M}})$ with a suitable topology such that $\alpha_w$ is a homeomorphism and such that ${\mathcal{F}}_0({\mathcal{M}})[[\hbar]]$ is sequentially dense in
$\mathcal{A}({\mathcal{M}})$. 

\subsection{Interacting models}\label{K_interaction}
In order to treat other interactions we introduce a new product $\cdot_T$ on ${\mathcal{F}}_{{\mathcal{M}}}$, the time ordered product. It is a commutative product which coincides with the $\ast$-product if the factors are time ordered, 
\begin{equation}
F\cdot_T G=F\ast G \text{ if } {\mathrm{supp}}(F)\gtrsim{\mathrm{supp}}(G) 
\end{equation}
where $\gtrsim$ means that there is a Cauchy surface such that the left hand side and the right hand side are in the future and past of the surface, respectively.
For the free field, we find
\begin{equation}
{\varphi}(f)\cdot_T {\varphi}(g)={\varphi}(f){\varphi}(g)+i\hbar \langle f,\Delta^D g\rangle
\end{equation}
with the "Dirac propagator" (see \cite{K-Dirac})
\begin{equation}
\Delta^D=\frac12(\Delta^R+\Delta^A) \ .
\end{equation}
It may be generalized to all of ${\mathcal{F}}_0({\mathcal{M}})[[\hbar]]$ by
\begin{equation}
(F\cdot_T G)({\varphi})=\sum_n \frac{i^n\hbar^n}{n!}\langle F^{(n)},(\Delta^D)^{\otimes n}G^{(n)}\rangle \ .
\end{equation}
In text books on quantum field theory, the time ordered product is usually defined for fields in the Fock space representation. But there the ideal generated by the field equation vanishes which is in contradiction to the fact that the Dirac propagator is not a solution of the homogenous Klein-Gordon equation. Thus the time ordering on Fock space is not well-defined as a product of operators. On ${\mathcal{F}}_0({\mathcal{M}})[[\hbar]]$, however, it is well defined and is even equivalent to the pointwise (classical) product $\cdot$. Namely, we introduce the "time ordering operator"
\begin{equation}
TF({\varphi})=\sum_n\frac{i^n\hbar^n}{n!}\langle (\Delta^D)^{\otimes n},F^{(2n)}({\varphi})\rangle 
\ .
\end{equation}
$T$ is a linear isomorphism, and
\begin{equation}
F\cdot_T G=T(T^{-1}(F)\cdot T^{-1}(G)) \ .
\end{equation}
In terms of $T$, explicit formulae for interacting fields can be given in terms of the formal S-matrix which is just the exponential function computed via the time ordered product,
\begin{equation}
S(F)=T \exp(T^{-1}(F)) \ .
\end{equation}
In terms of $S$ we can write down the analogue of the M\o{}ller operators for quantum field theory,
via Bogoliubov's formula
\begin{equation}
\left. F_V\equiv R_V(F)\equiv R(V,F)\doteq\frac{d}{d\lambda}S(V)^{-1}\star S(V+\lambda F)\right |_{\lambda=0}=
 S(V)^{-1}\star (S(V)\cdot_T F)
\end{equation}
where the inverse is built with respect to the $\star$-product. The interacting field $R_V$ is a linear map from ${\mathcal{F}}_0({\mathcal{M}})[[\hbar]]$ to itself and  describes the transition from the free action to the action with additional interaction term $V$. It satisfies two important conditions, retardation and equation of motion. 
As far as the retardation property is concerned, one observes that if ${\mathrm{supp}}(V)$ is causally later than ${\mathrm{supp}}(F)$, i.e. there exists a Cauchy surface that separates the supports, the time ordering and star products coincide, hence by associativity of both $R_V(F)=F$.
We now show that these interacting fields satisfy the off-shell field equation
\begin{equation}
R_V(\varphi(Kf))=\varphi(Kf)+i\hbar\,R_V(\langle V^{(1)},f\rangle)\ ,
\end{equation}
where $f\in {\cal D}(M)$ and $K$ is the Klein-Gordon operator.

Using $S(V)^{(1)}=S(V)\cdot_T V^{(1)}$ we obtain
\begin{align*}
R_V(\varphi(Kf))&= S(V)^{-1}\star (S(V)\cdot_T \varphi(Kf))\notag\\
&=S(V)^{-1}\star \Bigl(S(V)\cdot \varphi(Kf)+i\hbar\,S(V)\cdot_T \langle V^{(1)},\Delta_D K\, 
f\rangle \Bigr)\ .
\end{align*}
Due to $\Delta_D K=\mathrm{id}$, it remains to show that 
$S(V)^{-1}\star \Bigl(S(V)\cdot \varphi(Kf)\Bigr)
=\varphi(Kf)$. This holds true, since the contractions of $S(V)^{-1}$ with
$\varphi(Kf)$ vanish due to $\Delta K=0$, that is \[S(V)^{-1}\star \Bigl(S(V)\cdot \varphi(Kf)\Bigr)
=\Bigl(S(V)^{-1}\star S(V)\Bigr)\cdot \varphi(Kf)\ .\]

\subsection{Renormalization}\label{K_renormalization}
The remaining problem is the extension of the time ordered product to local functionals. Here the problem cannot be solved by the transition to an equivalent product
\begin{equation}
F\cdot_{T_H}G=\alpha_H (\alpha_H^{-1}(F)\cdot_T\alpha_H^{-1}(G)) \ . 
\end{equation}
This would amount to replacing the Dirac propagator by the Feynman like propagator
$\Delta^{D}+iH$. The wave front set of $\Delta^{D}+iH$ is
\begin{align*}
{\mathrm{WF}}(\Delta^{D}+iH)=\{(x,y,k,k')\in{\mathrm{WF}}(\Delta),k\in\overline{V_\pm}\text{ if }x &\in J_{\pm}(y) \}\\ &\cup\{(x,x,k,-k),k\not=0\} \ .
\end{align*}
One observes that pointwise products of these propagators exist outside of the diagonal.
The technical problem which has to be solved in renormalization is therefore to extend a distribution which is defined on the complement of some submanifold to the full manifold \cite{K-BF}.

The construction can be much simplified by the insight that the time ordered product coincides with the product $\ast$ for time ordered supports. For local functionals the time ordered product is therefore defined whenever the localizations are different. Namely, let $\mathcal{L}_i$, $i=1,\ldots,n$ be Lagrangians. Then the time ordered product can be defined as a  ${\mathcal{F}}$-valued distribution on ${\mathcal{M}}^n\setminus D$ where $D$ is the subset where at least two variables coincide. Moreover, once products of less than $n$ factors are everywhere defined, one can define the $n$-th order product outside of $\Delta_n$. 

The remaining problem is the extension of distributions from ${\mathcal{M}}^n\setminus \Delta_n$ to 
${\mathcal{M}}^n$. This can be done, but the process is not necessarily unique, in the following way: for simplicity we stick to the case of polynomial interactions and
we use the fact recalled above that for two (polynomial) Lagrangians $\mathcal{L}_1,\mathcal{L}_2\in\mathcal{F}_0(\mathcal{M})$ with $\mathrm{supp}(\mathcal{L}_1)$ later than $\mathrm{supp}(\mathcal{L}_2)$ the time ordered product coincides with the $\star$-product
\begin{equation}\label{time ordering}
\mathcal{L}_1\cdot_T \mathcal{L}_2 = \mathcal{L}_1\star \mathcal{L}_2 \ .
\end{equation}
This implies the following causality property of the $S$-matrix
\begin{description}
\item[\emph{Causality}.] $S(\mathcal{L}_1+\mathcal{L}_2)=S(\mathcal{L}_1)\star S(\mathcal{L}_2)\ $. 
\end{description}
The causality property determines the derivatives $S^{(n)}$ of $S$ at the origin 
\[\left. S^{(n)}(0)(\mathcal{L}^{\otimes n})\equiv S^{(n)}(\mathcal{L}^{\otimes n})\equiv\frac{d^n}{d\lambda^n}S(\lambda \mathcal{L})\right |_{\lambda=0} \ ,\]
(i.e.~the higher order time ordered products) partially in terms of lower order derivatives 
namely
\begin{equation}
S^{(n)}(\mathcal{L}_1^{\otimes k}\otimes \mathcal{L}_2^{\otimes (n-k)})= S^{(k)}(\mathcal{L}_1^{\otimes k})\star S^{(n-k)}(\mathcal{L}_2^{\otimes (n-k)})\ .
\end{equation}
While on $\mathcal{F}_0(\mathcal{M})$ this is an immediate consequence of the definition of the $S$-matrix and of time ordering, 
it is the key property by which an extension to local functionals can be made. Namely, local functionals can be splitted into a sum of terms which are localized in smaller regions. Together with the multilinearity of the higher derivatives this allows the determination of the $n$th order in terms of the derivatives with order less than $n$
for all elements of the tensor product $\mathcal{F}_{loc}(\mathcal{M})^{\otimes n}$ whose support is disjoint from the thin diagonal.
Here the support of $\sum(\mathcal{L}_1\otimes \cdots \otimes \mathcal{L}_n)$ is defined as the union of the cartesian products of the supports of Lagrangians $\mathcal{L}_k$.   
Together with the property 
\begin{description}
\item[\emph{Starting element}.] $S(0)=1$, $S^{(1)}=\mathrm{id}\ $,
\end{description}
this fixes the higher derivatives of $S$ at the origin partially on local functionals. 

The $\star$-product and the time ordered product $\cdot_T$ on $\mathcal{F}_0(\mathcal{M})$ were defined in terms of 
functional differential operators. Therefore the $S$-matrix $S(V)$, $V\in\mathcal{F}_0(\mathcal{M})$ at the field configuration $\varphi$ depends on $\varphi$ only via the functional derivatives of $V$ at $\varphi$. One then  requires that a similar condition holds true also for the extension of $S$ to $\mathcal{F}_{loc}(\mathcal{M})$.

A convenient additional condition is that, loosely speaking, $S$ should have no explicit dependence on $\varphi$,
\begin{description}
\item[\emph{Field Independence}.]  $\delta S/\delta\varphi=0\ $. 
\end{description}
For the action on $\mathcal{F}_0(\mathcal{M})$ this is the case due to the fact that the differential operators in terms of which time ordering, $\star$-product and topology were defined do not depend on $\varphi$. 

These conditions, supplemented by some other conditions on smoothness requirement w.r.t. the parameters of the theory (see, e.g., \cite{K-HW1,K-HW2,K-HW3,K-HW4}), suffice to make the extension to the full space of local interactions possible, but as recalled the extension is not uniquely determined. Moreover, as shown by Hollands and Wald \cite{K-HW1, K-HW2}, although in a slightly different framework, the algebras of renormalized interacting fields can be expressed in a functorial way as indicated in Section~\ref{K_ALCT}. That the time-slice axiom is also satisfied, in the Wick polynomials as well as interacting cases, is due to a recent investigation \cite{K-CF}.

The nonuniqueness is described in the following main theorem of renormalization:
\begin{theorem}
Let $S_i$ be extensions of the formal S-matrix to ${\mathcal{F}}_{\mathrm{loc}}({\mathcal{M}})$ with 
$S_i(F+G)=S_i(F)\ast S_i(G)$ if ${\mathrm{supp}}{F}\gtrsim{\mathrm{supp}}{G}$. Then there exist a formal diffeomorphism (tangent to the identity) $Z$ on ${\mathcal{F}}_{\mathrm{loc}}({\mathcal{M}})$ such that
\begin{equation}
S_2=S_1\circ Z \ .
\end{equation}
        
\end{theorem}

In other renormalization schemes, for instance in the Wilson-Polchinski framework of Flow Equations (see, e.g., \cite{K-Salmhofer}), one uses a regularized time ordered product $T_{\Lambda}$ which can directly be defined on ${\mathcal{F}}_{\mathrm{loc}}({\mathcal{M}})$. $\Lambda$ may be understood as a momentum cutoff (a concept with problems on a generic spacetime) or as another parameter which modifies the propagator like dimensional or analytic regularization. One thus has a direct definition of the regularized time ordered exponential $S_{\Lambda}$. It now follows from the Epstein-Glaser theory \cite{K-BDF1} that there exists a choice of formal diffeomorphisms $Z_{\Lambda}$ such that
\begin{equation}
S=\lim S_{\Lambda}\circ Z_{\Lambda} \ .
\end{equation}
Moreover, from the main theorem of renormalization, one finds that the choice of $Z_{\Lambda}$ (the ``counter terms") is unique up to a convergent family of formal diffeomorphisms. In particular, if one finds a regularization such that $S_{\Lambda}$ is meromorphic in $\Lambda$ with the origin representing the removal of regularization, then the principal part of the Laurent series can be used as a counter term, which can deviate from any allowed choice of counter terms only by a converging contribution. Thus 
the scheme of minimal subtraction which consists in subtracting the terms with negative powers  in $\Lambda$, is a possible choice of renormalization. It depends, however, on the choice of the regularization, and may be in conflict with other physical principles. This happens for instance with minimal subtraction in dimensional regularization if one wants to have supersymmetry.


\printindex


\begin{thebibliography}{99}

\bibitem{K-BF} R. Brunetti, K. Fredenhagen,
``Microlocal analysis and interacting quantum field 
theories: Renormalization on physical backgrounds,'' {\it Commun. Math. 
Phys.} {\bf 208} (2000) 623

\bibitem{K-BF2} R. Brunetti, K. Fredenhagen, work in progress.

\bibitem{K-BDF1} R. Brunetti, M. D\"utsch, K. Fredenhagen, ``Perturbative Algebraic Quantum Field Theory and the Renormalization Groups," Preprint in preparation (2008).


\bibitem{K-BFK} R. Brunetti, K. Fredenhagen, M. K\"ohler, ``The microlocal spectrum condition and Wick polynomials of free fields on curved spacetimes,'' {\it Commun. Math. Phys.} {\bf 180} (1996) 633--652

\bibitem{K-BFR} R. Brunetti, K. Fredenhagen, P. L. Ribeiro, work in progress.

\bibitem{K-BFV} R. Brunetti, K. Fredenhagen, R. Verch, ``The 
generally covariant locality principle -- A new paradigm for local 
quantum physics,'' {\it Commun. Math. Phys.} {\bf 237} (2003) 31

\bibitem{K-BOR} D. Buchholz, I. Ojima, H. Roos, ``Thermodynamic properties of non-equilibrium states in quantum field theory,'' {\it Ann. Phys. (N. Y.)} {\bf 297} (2002) 219

\bibitem{K-CF} B. Chilian, K. Fredenhagen, ``The time slice axiom in perturbative quantum field theory on globally hyperbolic spacetimes,"
eprint, arXiv:0802.1642 

\bibitem{K-Dirac} P. A. M. Dirac, ``Classical theory of radiating electrons,'' {\it Proc. Roy. Soc. of London} {\bf A929}, (1938) 148

\bibitem{K-Gin} N. Ginoux, ``Linear wave equations,'' in: Proceedings of ``Quantenfeldtheorie auf gekrŸmmten Raumzeiten''
UniversitŠt Potsdam, 8.-12.10.2007, available at http://users.math.uni-potsdam.de/~baer/QFT/Proceedings/qftkurs-proceedings.html

\bibitem{K-HK} R. Haag, D. Kastler, ``An algebraic approach to quantum field theory,"
{\it J. Math. Phys.} {\bf 5} (1964) 848

\bibitem{K-Haag} R. Haag, {\it Local Quantum Physics}, Berlin, Heidelberg, New-york, Springer-verlag, 1996, 2nd ed.

\bibitem{K-Hamilton} R. S. Hamilton, ``The inverse function theorem of Nash and Moser,'' {\it Bullettin (New Series) of the American Mathematical Society}, {\bf 7} (1982) 65

\bibitem{K-HW1} S. Hollands, R. M. Wald, ``Local Wick Polynomials
and Time-Ordered-Products of Quantum Fields in Curved Spacetime,''
{\it Commun. Math. Phys.}  {\bf 223} (2001) 289

\bibitem{K-HW2} S. Hollands, R. M. Wald, ``Existence of Local Covariant
Time-Ordered-Products of Quantum Fields in Curved Spacetime,''
{\it Commun. Math. Phys.} {\bf 231} (2002) 309

\bibitem{K-HW3} S. Hollands, R. M. Wald, ``On the Renormalization Group
  in Curved Spacetime,'' {\it Commun. Math. Phys.} {\bf 237} (2003) 123

\bibitem{K-HW4} S. Hollands, R. M. Wald, ``Conservation of the stress tensor
in interacting quantum field theory in curved spacetimes,''  {\it Rev. Math. Phys.} {\bf 17}  (2005) 227


\bibitem{K-Marolf} D. M. Marolf, ``The generalized Peierls brackets," {\it Ann. Phys.} {\bf 236} (1994) 392

\bibitem{K-Peierls} R. Peierls, ``The commutation laws of relativistic field theory," {\it Porc. Roy. Soc. (London)} {\bf A 214} (1952) 143

\bibitem{K-Peskin} M. E. Peskin, D. V. Schr\"oder, {\it An Introduction to Quantum Field Theory}, Perseus 1995. 

\bibitem{K-Radman} M. Radzikowski, ``Micro-local approach to the Hadamard condition in quantum field theory in curved spacetime," {\it Commun. Math. Phys} {\bf 179} (1996) 529

\bibitem{K-Salmhofer} M. Salmhofer, {\it Renormalization. An introduction.} Texts and Monographs in Physics. Springer-Verlag, Berlin, (1999)

\bibitem{K-St} A. Strohmaier, ``Microlocal Analysis,'' in: Proceedings of ``Quantenfeldtheorie auf gekrŸmmten Raumzeiten''
UniversitŠt Potsdam, 8.-12.10.2007, available at http://users.math.uni-potsdam.de/~baer/QFT/Proceedings/qftkurs-proceedings.html

\bibitem{K-de Witt} B. S. deWitt,``The spacetime approach to quantum field theory," in ``Relativity, Groups and Topology II: Les Houches 1983" (B. S. deWitt, R. Stora, eds.), part 2, North-Holland, New York (1984) 381

\end{thebibliography}
\end{document}